\documentclass[12pt, epsf]{article}
\usepackage{epsf}
\usepackage{graphicx}
\begin{document}
\newcommand {\sheptitle}
{ Degenarate neutrino mass models revisited                       }
\newcommand{\shepauthor}
{N. Nimai  Singh $^{\dag}$\footnote{Regular Associate of
    The Abdus Salam ICTP, Trieste, Italy. \\{\it{E-mail}}: nimai03@yahoo.com}, H. Zeen
  Devi$^{\dag}$     and  Abhijit Borah$^{\ddag}$,
   S. Somorendro Singh$^{*}$} 
\newcommand{\shepaddress}
{$^{\dag}$ Department of Physics, Gauhati University, Guwahati-781 014,India \\
$^{\ddag}$ Department of Physics, Fazl Ali College, Mokokchung- 798 601, India\\
$^{*}$Department of Physics and Astrophysics, University of Delhi, Delhi- 110 007, India  }
\newcommand{\shepabstract}
 {A  parametrisation of the degenerate neutrino mass matrix  obeying $\mu -\tau$ symmetry, is introduced for  detailed numerical analysis. The present
 parametrisation for degenerate models has the ability to  lower the solar
 mixing angle below the tri-bimaximal 
value $\tan^{2}\theta_{12}=0.5$,  while maintaining the condition
 of maximal atmospheric mixing angle and zero reactor angle.
  The combined  data on the mass-squared
 differences derived from various  oscillation experiments, and also
 from the  bounds on absolute neutrino masses in $0\nu\beta\beta$
 decay and cosmology, gives certain constraints on the validity of the
 degenerate models. \\
{\bf Keywords}. Degenerate neutrino mass model, absolute neutrino masses,
 solar mixing angle below tribimaximal mixings.
\\
{\bf PACS} Nos 14.60.Pq; 12.15.Ff; 13.40.Em     
 } 

\begin{titlepage}
\begin{flushright}
\end{flushright}
\begin{center}
{\large{\bf\sheptitle}}
\bigskip \\
\shepauthor
\\
\mbox{}
\\
{\it \shepaddress}
\\
\vspace{.5in}
{\bf Abstract} \bigskip \end{center}\setcounter{page}{0}
\shepabstract
\end{titlepage}

\section{Introduction}
Discrimination of neutrino mass patterns among three
possible cases, viz., degenerate, normal and inverted hierarchical
models, has drawn considerable attentions in the last one decade. This
has renewed interest with the presently available precise observational data
from neutrino oscillation experiments as well as bounds on absolute
neutrino masses. The present neutrino oscillation data gives the following
 information  for the  parameters at 3$\sigma$ from global $3\nu$ oscillation analysis.\cite{ref SNO}
\begin{center}
$\bigtriangleup m^2_{21}(10^{-5}eV^2)=(7.4-7.9)$ ( best-fit at $7.7$),\\ 
$|\bigtriangleup m^2_{32}(10^{-3}eV^2)|=(2.1-2.8)$ ( best-fit at $2.4$),\\
 $\tan^{2}\theta_{12}=(0.41-0.58)$ (best-fit at $0.45$),\\ 
$\tan^{2}\theta_{23}=(0.56-2.03)$ ( best-fit at $1.0$),\\ 
$\sin^{2}\theta_{13}=0.016\pm0.010$. 
\end{center}
At present  there is no precise information about the absolute
neutrino masses. One could  expect only the upper bounds on the absolute neutrino masses
 from various experiments like the Tritium $\beta$ decay, Neutrinoless double beta
 decay and the cosmological observations. The bounds on the absolute
 neutrino masses are \cite{ref PP,ref WMAP}
\begin{center}
$m_{\nu_{e}}=(\sum_{i} m_{i}^{2}|U_{ei}|^{2})^{1/2}< 2.3 eV$ (Tritium
  $\beta$ decay) \\
$m_{ee}=|\sum_{i}m_{i}U_{ei}^{2}| < 0.3-1.0 eV$ ($0\nu\beta\beta$ decay)\\
$m_{cosmo}=\sum_{i}|m_{i}|<0.61 eV $ (cosmological bounds)
\end{center} 
While both normal and inverted hierarchical models are well within the
above observational bounds, and are therefore far from ruling out
at the moment, the degenerate models are considered almost disfavoured
in the literature,
following the present bounds on absolute neutrino mases, particularly
from cosmological and neutrinoless double beta decay bounds. However a
closer analysis reveals that such assertion is only  correct, specifically for
larger absolute neutrino masses $m_{0}\sim 0.4 eV$. There are still
 possibilities for certain degenerate models which allow lower values
of neutrino masses $m_{0}\sim 0.1 eV$ valid for lower values of solar
mixings. This is one of the objectives for further investigations in the
present work.

Degenerate models are generally classified according to their
CP-parity pattern in their mass eigenvalues $m_i=(m_1, m_2, m_3)$,
viz., Type IA: (+-+), Type IB:(+++), Type IC: (++-) respectively. It
will be shown in the present work that both Type IB and IC are almost
disfavoured by the presently available data on absolute neutrino
masses. This is due to the fact that the predicted  absolute neutrino
masses are in the range $m_0\sim 0.4 eV$ in these two cases. On the
other hand, Type IA has many
 interesting properties which are yet to be explored, particularly variation of $m_0$
with solar mixing $\tan^2\theta_{12}$, and also a partial cancellation of
even and odd CP parity in the first two mass eigenvalues appeared in
the expressions of $m_{\nu_{e}}$ and $m_{ee}$. This makes
Type IA degenerate model far from ruling out. It provides
enough scope for future experiments to go  the sensitivity down to
$|m_{ee}|=0.03 eV$. 

In the theoretical front there are several attempts to find out the
most viable neutrino mass models, and among them the neutrino mass models
obeying $\mu-\tau$ symmetry \cite{ref INCMPLTE,ref AKKL,ref KOI,ref KT,ref MNY}, have drawn considerable attention.
 Neutrino mass matrix having $\mu-\tau$ symmetry, leads to maximal
 atmospheric mixings ($\tan^{2}\theta_{23}=1$) and
 zero reactor angle($\theta_{13}=0$), whereas the solar angle  is purely
 arbitrary. This has to be fixed by the input values of the  parameters in the
 mass matrix. The tribimaximal mixings(TBM)\cite{ref HPS,ref EMA} with
 $\tan^2\theta_{12}=0.5$ is a special case of this symmetry.
 There are four unknown elements present in a general $\mu-\tau$
 symmetric mass matrix and  it is difficult to solve these four
 unknown elements from only three equations  involving
 observational data on  $\tan 2\theta_{12}$, $\bigtriangleup m^2_{21}$ and
 $\bigtriangleup m^2_{32}$. Thus
\begin{center}
\begin{equation}\label{ch501}
 m_{LL}=
 \left(\begin{array}{ccc}
 m_{11}& m_{12} & m_{12} \\
  m_{12} & m_{22}  &  m_{23}  \\
 m_{12} & m_{23}  & m_{22}
 \end{array}\right)
 \end{equation}
\end{center}
 where the eigenvalues and solar mixings are
$$m_1=m_{11}-\sqrt{2}\tan\theta_{12} m_{12},$$
$$m_2=m_{11}+\sqrt{2}\cot\theta_{12} m_{12},$$
$$m_3=m_{22}-m_{23},$$
$$\tan 2\theta_{12}=\frac{2\sqrt{2}m_{12}}{m_{11}-m_{22}-m_{23}}.$$
In our earlier works we simply considered possible parametrisation with
lesser numbers of free parameters consistent with available data for
practical solution of the mass matrix. We
have already reported our method for parametrisation with only two parameters $\eta$
and $\epsilon$ and  the ratio of the these two parameters named as flavour 
twister \cite{ref NMB}, is responsible for lowering the solar
angle below tri-bimaximal solar mixing.
 We have already parametrised both  normal and inverted
 hierarchical
 neutrino mass matrices\cite{ref NZM}. In TBM mixing we have the value
 of the solar mixing angle,($\tan^{2}\theta_{12}=0.5$). But, the
 recent
 global 3$\nu$ oscillation analysis has shown a mild
 departure from the tribimaximal neutrino mixings,
 $\tan^2\theta_{12}=0.45$ and the present work has its own relevance
 here. Such parametrisation thus reduces to two
 unknown free parameters $\eta$ and $\epsilon$ in addition to an
 overall neutrino mass scale $m_0$, making the three equations
 solvable in practice.

In section 2  we first extend our earlier method of parametrisation of
$\mu-\tau$ symmetric mass matrices, to  a particular degenerate neutrino mass model
 Type 1A($m_{1},-m_{2},m_{3}$) denoted by CP-parity pattern (+-+), and
 also identify
 the flavour twister term  to obtain tribimaximal
 mixings and then modify it for deviations from tribimaximal
 mixings. Similar analysis is also carried out for Type IB and IC
 degenerate models.Section 3 concludes with a brief summary.

\section{Parametrisation of degenerate models}

\subsection{Degenerate Type 1A $(m_1, -m_2, m_3$)}

The zeroth order left-handed Majorana mass matrix \cite{ref AF} for
Type IA degenerate model, is given by 
\begin{center}
\begin{equation}\label{ch501}
 m_{LL}^{o}=
 \left(\begin{array}{ccc}
 0&-1/\sqrt{2}& -1/\sqrt{2} \\
  -1/\sqrt{2}&  1/2  &   -1/2  \\
 -1/\sqrt{2}&  -1/2  &  1/2
 \end{array}\right)m_0
 \end{equation}
\end{center}
where $m_{0}$ is the overall scale factor and mass eigenvalues are
 $Diag(1,-1,1) m_{0} $.
 A complete  light neutrino mass matrix $m_{LL}$ for the degenerate neutrino
 mass model (Type 1A) with $\mu-\tau$ symmetry in eq.(2) is now modified as 
\begin{center}
\begin{equation}\label{ch502}
 m_{LL}=
 \left(\begin{array}{ccc}
 \epsilon- 2\eta  &   -c \epsilon  &   -c \epsilon \\
  -c\epsilon   &  1/2- d\eta  &   -1/2-\eta  \\
 -c\epsilon  &  -1/2-\eta  &  1/2-d\eta
 \end{array}\right)m_0
 \end{equation}
\end{center}
where $c$ and $d$ are just real constants, and $\eta$ and $\epsilon$ are
the unknown parameters. The mass matrix still maintains the $\mu-\tau$ symmetry but the
 arbitrary solar 
angle can  be fixed at the particular value by choosing   a set of values for 
 c,d, $\eta$ and $\epsilon$.

The mass matrix in eq.(3) predicts an  expression for solar mixing angle,
\begin{center} 
\begin{equation}\label{ch503}
 \tan 2\theta_{12}=\frac{2\sqrt{2}(-c\epsilon) }{\epsilon-2\eta-1/2+d \eta+1/2+\eta}
=-\frac{2\sqrt{2}c}{1+(d-1)\eta/\epsilon}
\end{equation}
\end{center}
 In general, any mass matrix obeying $\mu-\tau$ symmetry, can be
 diagonalized by
 the following unitary matrix,
\begin{center}
\begin{equation}\label{ch503}
 U_{MNS}=
 \left(\begin{array}{ccc}
 \cos\theta_{12} &  -\sin\theta_{12}  &   0 \\
  \frac{\sin\theta_{12}}{\sqrt 2}   &  \frac{\cos\theta_{12}}{\sqrt 2}    &-\frac{1}{\sqrt 2} \\
 \frac{\sin\theta_{12}}{\sqrt 2} &  \frac{\cos\theta_{12}}{\sqrt 2}  & \frac{1}{\sqrt 2}\\
 \end{array}\right)
 \end{equation}
\end{center}
Diagonalizing the mass matrix (3) we get the three mass eigenvalues as,
\begin{eqnarray}\label{ch504}
m_{1}=\frac{1}{2}(\epsilon-3\eta-d \eta+ \chi)m_{0},\\
m_{2}=\frac{1}{2}(\epsilon-3\eta-d \eta-\chi)m_{0},\\
m_{3}=(1+\eta-d\eta)m_{0},\\
\chi^2=\epsilon^2 +8c^{2}\epsilon^2 -2\eta\epsilon+2d
\eta\epsilon+\eta^{2}-2d \eta^{2} +d^{2}\eta^{2}\\
\end{eqnarray}
The choice of  $c=d=1$ in eq.(4) leads to 
tribimaximal mixings (TBM), $\tan^2\theta_{12}=0.5$ 
($\tan2\theta_{12}=2\sqrt{2})$. With the input value $m_0=0.4 eV$, the
values of the free  parameters are extracted from the data of neutrino
oscillation mass parameters, as
$\epsilon=0.661145$ and $\eta=0.165348$ respectively. As seen in (4)
there is no flavour twister term in this case.

 In the next step the flavour twister $\eta/\epsilon$ responsible for lowering solar angle,  is
then introduced by taking $c<1$ and $d>1$. Using the known values of
$\eta$ and $\epsilon$ already fixed in TBM, along with those of  c and d for a particular choice of
solar angle, the value of $m_0$ is extracted. Table 1 gives the 
 numerical results for a few selected cases. A scattered plot in
Fig.1 using the points generated within the allowed ranges of neutrino
oscillation mass parameters,  depicts the dependence of $\tan^2\theta_{12}$ on $m_0$.
This leads to  $m_{cosmos}$ within the upper bound from cosmology. For
$\tan^2\theta_{12}=0.45$,
we get $\sum|m_i|=0.275 eV$ and $|m_{ee}|=0.033 eV$. It can be
emphasised that the predictions on other oscillation parameters are
consistent with latest data.\\


\begin{table}[tbp]
\begin{tabular}{|l|l|l|l|} \hline
Parameters & I & II &  III \\ \hline
c          & 1.0      & 0.931 &  0.868 \\
d          & 1.0      & 1.011 & 1.025 \\
$m_{0}(eV)$      & 0.4      & 0.141 &  0.100 \\
\hline
$m_{1}(eV)$      &0.39664  & 0.13119 &  0.08749 \\
$m_{2}(eV)$      &-0.39674 & -0.13148 &  -0.08793\\
$m_{3}(eV)$      &0.40000       & 0.14074 &  0.09959\\ 
\hline
$\bigtriangleup m^2_{21}(10^{-5}eV^2)$& 7.838 & 7.651 & 7.685\\
$\bigtriangleup m^3_{23}(10^{-3}eV^2)$& 2.600 & 2.521 &  2.186 \\
$\tan^{2}\theta_{12}$ & 0.5 & 0.475 & 0.450 \\
$|m_{ee}|(eV) $& 0.132 & 0.047 & 0.033\\
$\sum|m_{i}|(eV) $ & 1.193 & 0.403 &  0.275  \\
 \hline
\end{tabular}
\hfil
\caption{\footnotesize  Numerical calculation for Type IA degenerate model. 
Different values of parameters c,d,$\epsilon$ and $\eta$
 along with the corresponding ranges of $\bigtriangleup m^2_{21}$ 
and$\bigtriangleup m^2_{23}$, $m_{ee}$,$\sum|m_{i}|$ and $\tan^{2}\theta_{12}$ for fixed value of
$\tan^{2}\theta_{23}=1.0$ and $\sin\theta_{13}=0$ }
\end{table}

\begin{figure}[!h]
\begin{center}
\includegraphics[width=12cm]{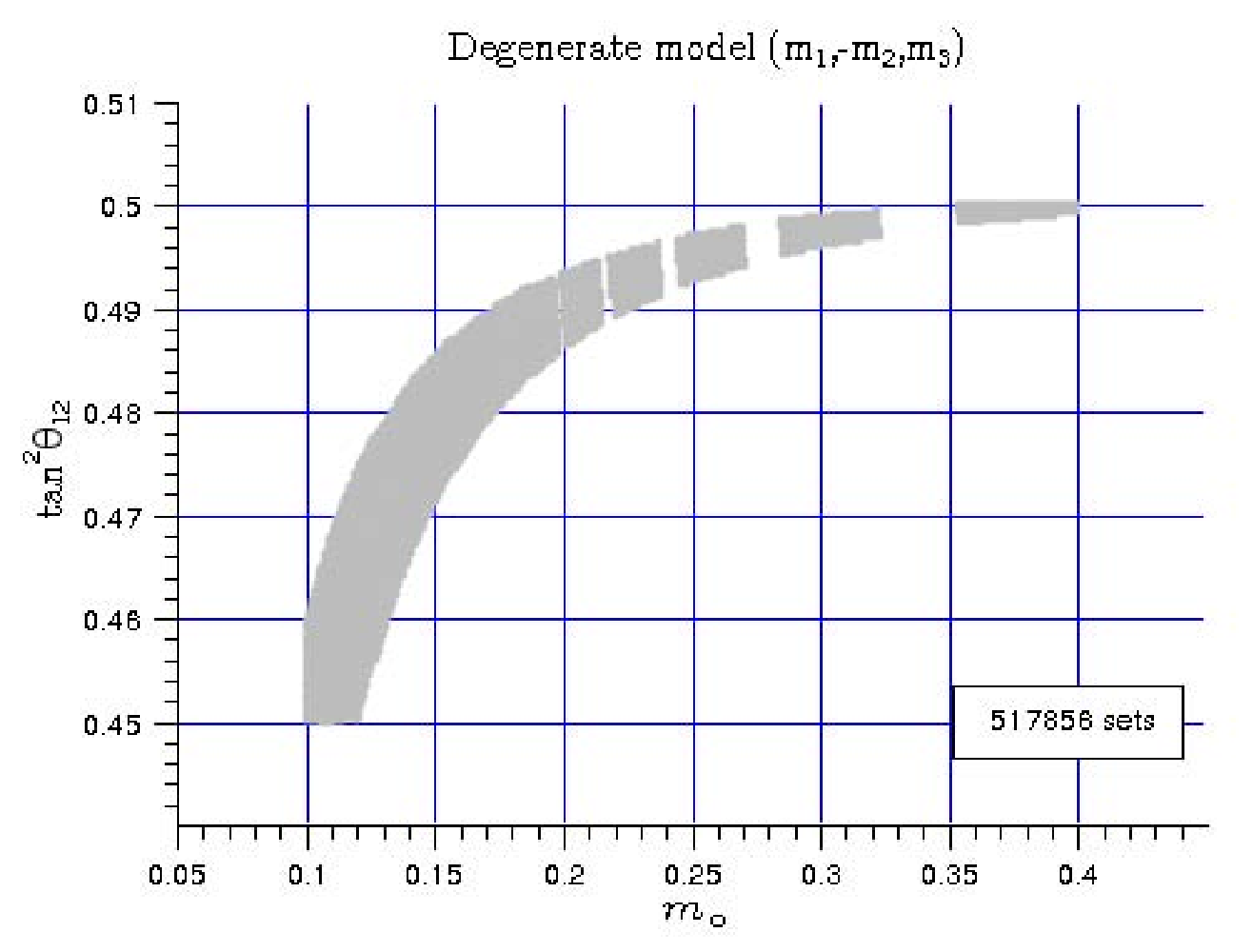}
\caption{Scattered plot showing the variation of solar mixing
  $\tan^2\theta_{12}$ with the scale of absolute neutrino masses $m_0(eV)$
in Type IA degenerate model with CP parity patern (+-+)}
\end{center}
\end{figure}

\begin{figure}[!h]
\begin{center}
\includegraphics[width=12cm]{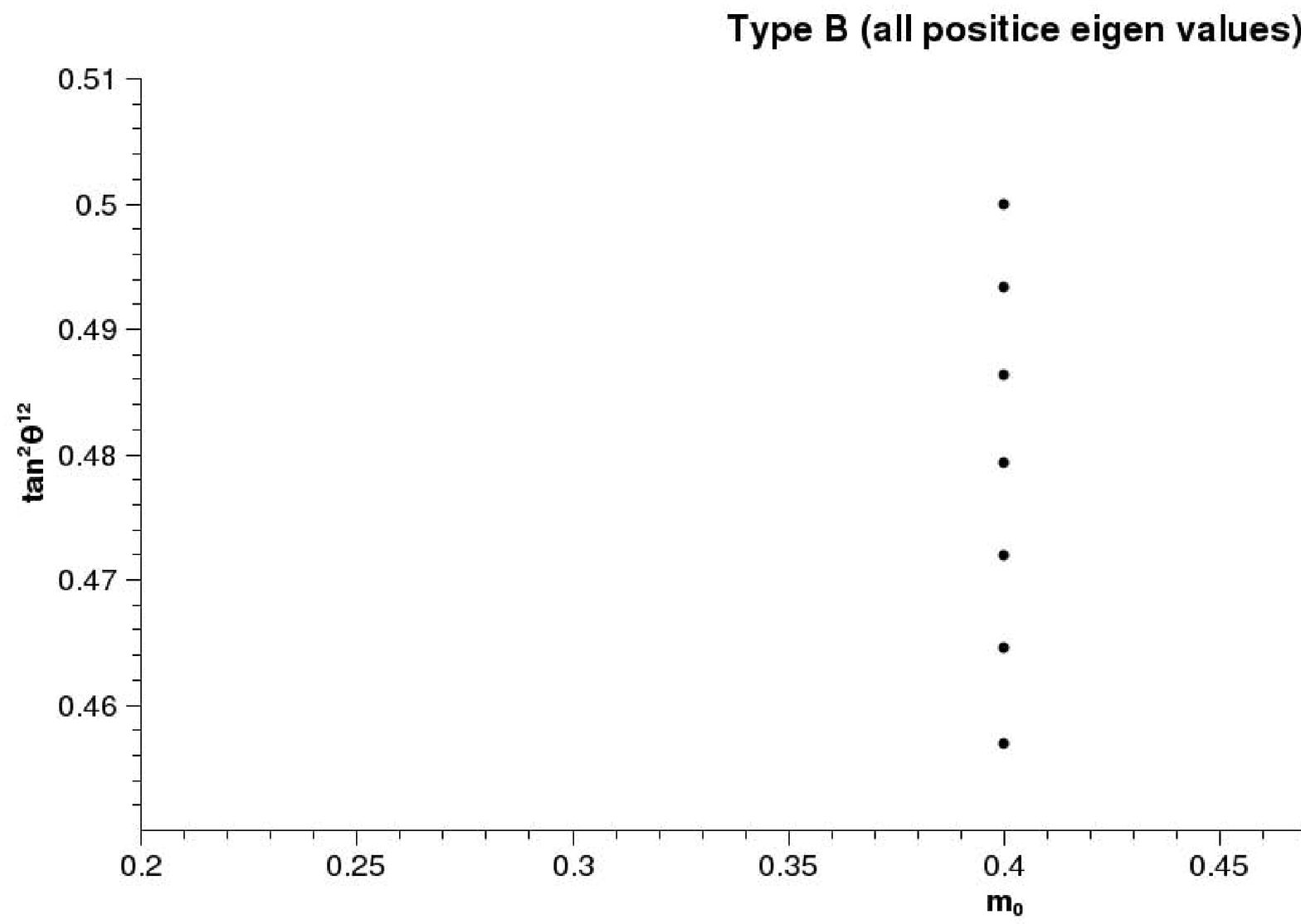}
\caption{  Scattered plot showing a relation of solar mixing
  $\tan^2\theta_{12}$ with the scale of absolute neutrino masses $m_0 (eV)$
in Type IB degenerate model with CP parity patern (+++)}
\end{center}
\end{figure}


\subsection{Degenerate Types 1B and 1C}

We now briefly perform similar analysis for other two degenerate models - Type
IB with CP-Parity (+++) and Type IC with (++-). The structures of their
mass matrices are given below:

\underline{Degenerate Type 1B($m_{1},m_{2},m_{3}$)} 
\begin{center}
\begin{equation}\label{ch502}
 m_{LL}=
 \left(\begin{array}{ccc}
 1-\epsilon- 2\eta  &   -c \epsilon  &   -c \epsilon \\
  -c\epsilon   &  1- d\eta  &   -\eta  \\
 -c\epsilon  &  -\eta  &  1-d\eta
 \end{array}\right)m_0
 \end{equation}
\end{center}
where the zeroth-order mass matrix with mass eigenvalues
$Diag(1,1,1)m_0$, is given by 
\begin{center}
\begin{equation}\label{ch502}
 m^{0}_{LL}=
 \left(\begin{array}{ccc}
 1  &   0  &   0 \\
  0  &  1  &   0  \\
 0 &  0  &  1
 \end{array}\right)m_0
 \end{equation}
\end{center}

\underline{Degenerate Type 1C($m_{1},m_{2}, -m_{3}$)} 
\begin{center}
\begin{equation}\label{ch502}
 m_{LL}=
 \left(\begin{array}{ccc}
 1-\epsilon- 2\eta  &   -c \epsilon  &   -c \epsilon \\
  -c\epsilon   &  - \eta  &   1-d \eta  \\
 -c\epsilon  &  1-d\eta  &  -\eta
 \end{array}\right)m_0
 \end{equation}
\end{center}
where the zeroth-order mass matrix with mass eigenvalues
$Diag(1,1,-1)m_0$, is given by 
\begin{center}
\begin{equation}\label{ch502}
 m^{0}_{LL}=
 \left(\begin{array}{ccc}
 1  &   0  &   0 \\
  0  &  0  &   1  \\
 0 &  1  &  0
 \end{array}\right)m_0
 \end{equation}
\end{center}

The above two models (Type IB and IC) are similar except the  interchange of (22) and (23)
elements, which imparts a negative sign before the third mass
eigenvalue $m_3$. Both give the same  expression for the  solar mixing angle as
\begin{center} 
\begin{equation}
 \tan 2\theta_{12}
=\frac{2\sqrt{2}c}{1+(1-d)\eta/\epsilon}
\end{equation}
\end{center}
Using the condition $c=d=1$ for tribimaximal mixings and input value $m_0=0.4eV$,  the values of
$\eta$ and $\epsilon$ are solved as $\eta=8.3138\times 10^{-5}$ and
$\epsilon=0.00395$.
 These two values are again used for lowering 
the solar angle corresponding to the condition $c<1$ and $d<1$. However the neutrino mass
scale is always obtained at $m_0=0.4 eV$ leading to $\sum|m_i|=1.194eV$ and
$m_{ee}=0.397 eV$ respectively. Table 2 gives some representative
numerical examples. There is no variation of solar angle with neutrino mass as
shown in Fig.2.  Such degenerate models are generally disfavoured
by the bounds with absolute neutrino masses.
\begin{table}[tbp]
\begin{tabular}{|l|l|l|l|} \hline
Parameters & I & II & III  \\ \hline
c          & 1.0      & 0.945 & 0.868  \\
d          & 1.0      & 0.998 & 0.998  \\
$m_{0}(eV)$      & 0.4      & 0.4 & 0.4  \\
\hline
$m_{1}(eV)$      &0.39677  & 0.39678 & 0.39678  \\
$m_{2}(eV)$      &0.39687 & 0.39687 & 0.39687 \\
$m_{3}(eV)$      &0.4000       & 0.40000 & 0.40000 \\ 
\hline
$\bigtriangleup m^2_{21}(10^{-5}eV^2)$& 7.936 & 7.619 & 7.143 \\
$\bigtriangleup m^3_{23}(10^{-3}eV^2)$& 2.492 & 2.494 & 2.494\\
$\tan^{2}\theta_{12}$ & 0.5 & 0.45 & 0.421  \\
$|m_{ee}|(eV) $& 0.397 & 0.397 &0.397 \\
$\sum|m_{i}|(eV) $ & 1.194 & 1.194 & 1.194   \\
 \hline
\end{tabular}
\hfil
\caption{\footnotesize  Numerical calculation for Type IB degenerate model.
 Different values of parameters c,d,$\epsilon$ and $\eta$
 along with the corresponding ranges of $\bigtriangleup m^2_{21}$ 
and$\bigtriangleup m^2_{23}$, $m_{ee}$,$\sum|m_{i}|$ and $\tan^{2}\theta_{12}$ for fixed value of
$\tan^{2}\theta_{23}=1.0$ and $\sin\theta_{13}=0$ }
\end{table}


\section{Summary and conclusion}

To summarise, we have presented a method of parametrisation of 
the degenerate neutrino mass models  obeying $\mu -\tau$ symmetry, and
this enables  to  lower the solar mixing angle below the tri-bimaximal 
value $\tan^{2}\theta_{12}=0.5$,  while maintaining the conditions
 of maximal atmospheric mixing angle and zero reactor angle.
  The predictions from these mass models
 are consistent with the data on the mass-squared
 differences derived from various  oscillation experiments, and also
 from the  bounds on absolute neutrino masses from $0\nu\beta\beta$
 decay and cosmology. Type IA degenerate model with CP-parity pattern
 (+-+) predicts a  variation of absolute neutrino mass scale with the 
 solar mixing angle, and the best-fit  value $\tan^2\theta_{12}=0.45$
 corresponds to  the abolute neutrino mass scale $0.1 eV$.
 The  model  is is not yet ruled out and therefore far from discrimination. The other two types IB and
 IC having  CP parity patterns (+++)
 and (++-) respectively, do not possess such variation and  the
 neutrino mass scale is almost  fixed at about $0.4 eV$ for a wide range of solar
 mixing angle.  These models are found to be almost  disfavoured by the availabe data from
 absolute neutrino masses. The present result has profound
 implications
 for future experiments on the discrimination of neutrino
 mass hierarchy. 
\section*{Acknowledgement}
One of us HZD would like to thank CSIR for the Senior Research
Fellowship as financial assistance for carrying out her research work.

\end{document}